\documentclass{aa}
\usepackage{graphicx}
\usepackage{amsmath}
\usepackage[]{natbib}
\bibpunct{(}{)}{;}{a}{}{,}
\usepackage{txfonts}

\begin{document}
   \title{Hot exozodiacal dust resolved around Vega with IOTA/IONIC}
   \titlerunning{Hot exozodiacal dust resolved around Vega with IOTA/IONIC}

   \author{D. Defr\`ere\inst{1}
          \and
          O. Absil\inst{2}
          \and
          J.-C. Augereau\inst{3}
          \and
          E. di Folco\inst{4,6}
          \and
          J.-P. Berger\inst{5}
          \and
          V. Coud\'e du Foresto\inst{6}
          \and
          P. Kervella\inst{6}
          \and
          J.-B. Le Bouquin\inst{3}
          \and
          J. Lebreton\inst{3}
          \and
          R. Millan-Gabet\inst{7}
          \and
          J.D. Monnier\inst{8}
          \and
          J. Olofsson\inst{9}
          \and
          W. Traub\inst{10}
         }


   \institute{Max Planck Institut f\"{u}r Radioastronomie, Auf den H\"{u}gel 69, 53121 Bonn, Germany\\
              \email{ddefrere@mpifr-bonn.mpg.de}
         \and
             Dept. d'Astrophysique, G\'eophysique \& Oc\'eanographie, Universit\'e de Li\`ege, 17 All\'ee du Six Ao\^ut, B-4000 Li\`ege, Belgium
         \and
             IPAG-UMR 5274, CNRS and Universit\'e Joseph Fourier, BP 53, 38041 Grenoble, France
         \and
             Laboratoire AIM, CEA Saclay-Universit\'e Paris Diderot-CNRS, DSM/Irfu/Service d'Astrophysique, 91191 Gif-sur-Yvette, France
         \and
             European Southern Observatory, Alonso de Cordova, 3107, Vitacura, Chile
         \and
             LESIA, Observatoire de Paris, CNRS\,UMR\,8109, UPMC, Universit\'e Paris Diderot, 5 place Jules Janssen, 92195 Meudon, France
         \and
         		 NASA Exoplanet Science Institute (Caltech), MS 100-22, 770 South Wilson Avenue, Pasadena, CA 91125, USA
         \and
         		 Astronomy Department, University of Michigan, Ann Arbor, MI 48109, USA
         \and
         	  Max Planck Institut f\"{u}r Astronomie, Königstuhl 17, D-69117 Heidelberg, Germany
         \and
            Jet Propulsion Laboratory (NASA/JPL), MS 301-355, 4800 Oak Grove Drive, Pasadena, CA 91109, USA
             }

   \date{Received; accepted}

  \abstract
   {Although debris discs have been detected around a significant number of main-sequence stars, only a few of them are known to harbour hot dust in their inner part where terrestrial planets may have formed. Thanks to infrared interferometric observations, it is possible to obtain a direct measurement of these regions, which are of prime importance for preparing future exo-Earth characterisation missions.
   }
   {We resolve the exozodiacal dust disc around Vega with the help of infrared stellar interferometry and estimate the integrated H-band flux originating from the first few AUs of the debris disc.}
   {Precise H-band interferometric measurements were obtained on Vega with the 3-telescope IOTA/IONIC interferometer (Mount Hopkins, Arizona). Thorough modelling of both interferometric data (squared visibility and closure phase) and spectral energy distribution was performed to constrain the nature of the near-infrared excess emission.}
   {Resolved circumstellar emission within $\sim$6\,AU from Vega is identified at the 3-$\sigma$ level. The most straightforward scenario consists in a compact dust disc producing a thermal emission that is largely dominated by small grains located between 0.1 and 0.3\,AU from Vega and accounting for 1.23 $\pm$ 0.45\% of the near-infrared stellar flux for our best-fit model. This flux ratio is shown to vary slightly with the geometry of the model used to fit our interferometric data (variations within $\pm$0.19\%).}
   {The presence of hot exozodiacal dust in the vicinity of Vega, initially revealed by K-band CHARA/FLUOR observations, is confirmed by our H-band IOTA/IONIC measurements. Whereas the origin of the dust is still uncertain, its presence and the possible connection with the outer disc suggest that the Vega system is currently undergoing major dynamical perturbations.}


   \maketitle
%

\section{Introduction}
\label{sec:intro}

The discovery of a debris disc around Vega (HD 172167, A0V, 7.76\,pc) by \cite{Aumann:1984} was one of the first hints that extrasolar planetary systems exist. The debris disc was identified thanks to an infrared excess (beyond 12\,$\mu$m) with respect to the expected photospheric flux by the Infrared Astronomical Satellite (IRAS). During subsequent decades, a successive generation of instruments (e.g., ISO, SCUBA, Spitzer, AKARI, Herschel) confirmed its existence over a wide wavelength range covering the infrared and millimetre regimes. The appearance of the disc has been found to vary significantly across this wavelength domain. Sub-millimetre and millimetre observations revealed a clumpy structure of large dust grains located  between about 80 and 120\,AU  \citep[e.g.,][]{Holland:1998,Koerner:2001, Wilner:2002, Marsh:2006}.\footnote{These results are however challenged by more recent observations with the Plateau de Bure Interferometer \citep{Pietu:2011}.} If present, the clumps are thought to be dust grains trapped into mean motion resonances with a planet located near to the disc, as first proposed by \cite{Wilner:2002}, and modelled later by \cite{Wyatt:2006} and \cite{Reche:2008}. Infrared observations revealed a smooth axi-symmetric structure extending from 85 to at least to 815\,AU and containing about $3\times10^{-3}M_\oplus$ of dust grains \citep[e.g.,][]{Su:2005,Rieke:2005,Sibthorpe:2010}. This huge size of the disc seen in the infrared came as a surprise and raised several questions about the mechanism at the origin of the dust. \cite{Su:2005} suggest that the extended disc is the result of a recent massive collision of planetesimals and the subsequent collisional cascade. This would produce a high-mass disc composed of very small grains (less than a few $\mu$m), which are blown out of the system by radiation pressure immediately upon creation, resulting in the observed large disc extent. Given the age of Vega \citep[about 455\,Myr,][]{Yoon:2010}, the statistical likelihood of such an event occurring with two bodies of enough mass to explain the sub-millimetre observations is, however, quite low \citep{Wyatt:2002}. Conversely, \cite{Muller:2010} succeeded in reproducing the surface brightness radial profile using intermediate size grains in elliptical orbits around the parent planetesimal ring, and therefore conclude that it is consistent with a steady-state model. Finally, another interesting model that was able to reproduce the infrared observations suggests that the debris disc of Vega is the result of icy planet formation \citep{Kenyon:2008}.

These studies of the outer part of the debris disc around Vega have been complemented by infrared interferometric observations of the inner part. The first attempt was realised with the PTI interferometer in dispersed mode \citep{Ciardi:2001}. The 110-m long baseline and the poor spatial frequency coverage were, however, not well adapted to drawing a clear conclusion and a simple debris-disc model accounting for 3 to 6\% of Vega's K-band flux was proposed as the most likely scenario to explain the observations. A more recent study using the shortest baselines at the CHARA array has derived a K-band flux ratio between the stellar photosphere and the debris disc of 1.26 $\pm$ 0.28\% within the FLUOR field-of-view \citep[about 8\,AU in radius,][]{Absil:2006b,Absil:2007b}. In the N band, the best constraint on the thermal emission from warm dust was obtained by nulling interferometry, with no resolved emission above 2.1\% of the level of stellar photospheric emission at separations larger than 0.8\,AU \citep{Liu:2009}. The hot dust grains observed in the inner debris disc of Vega are believed to derive from collisions between larger rocky bodies and/or by the evaporation of comets, as in the solar zodiacal disc. The inferred dust populations are, however, much hotter, more massive, and composed of much smaller grains than the zodiacal cloud. Such grains would be expected to be expelled from the inner planetary system by radiation pressure within only a few years, which indicates inordinate replenishment rates.

\begin{figure}[!t]
\centering
\includegraphics[width=5.0 cm]{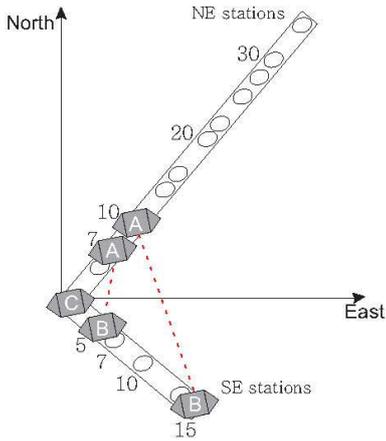}
\caption{Array geometry given with the available stations. Stations used during the observations are marked with a letter. Telescopes A and C can move on stations located along the 35\,m north-eastern arm, while telescope B can move along the 15\,m south-eastern arm.}
\label{fig:IOTA}
\end{figure}

The present paper reports on new infrared interferometric observations of the hot inner part of the debris disc around Vega. Visibility and closure phase H-band measurements were obtained in June 2006 with the 3-telescope IOTA/IONIC instrument, one year after the pioneering K-band observations of \cite{Absil:2006b}. Studying the inner debris disc around Vega is of prime importance as it corresponds to the location where terrestrial planets are supposed to be formed and evolve, and will provide additional constraints for discussing the still-debated origin of the dust. On a longer time scale, the characterisation of inner debris discs is also relevant for preparing the programme of future space missions dedicated to the direct detection and characterisation of Earth-like planets, since the presence of large quantities of warm dust in the habitable zone around nearby main sequence stars might jeopardize the success of such missions \citep{Roberge:2009,Defrere:2010}.

\section{Observations and data reduction}\label{sec:data}
\subsection{Instrumental setup}

The interferometric data discussed in this paper were obtained with the IONIC combiner at the IOTA interferometer \citep[Infrared-Optical Telescope Array,][]{Traub:2003}. IOTA was a three telescope interferometer located at the Fred Whipple Observatory atop Mount Hopkins (Arizona, USA). The IOTA telescopes were movable among several stations along an L-shaped track: telescopes A and C can move along the 35-m north-eastern arm, while telescope B moves along the 15-m south-eastern arm, see Figure~\ref{fig:IOTA}. This enables synthesizing an aperture of 35\,m $\times$ 15\,m, which corresponds to an angular resolution of about 5-12\,mas at 1.65\,$\mu$m. The collecting optics consisted of 0.45-m Cassegrain primary mirrors, which were fed by siderostats. Tip-tilt servo systems mounted behind the telescopes compensated for the atmospherically induced motion of the images, while delay lines actively tracked the fringes by adjusting the optical path delay between the different baselines. All the beams were deflected by a series of mirrors into the laboratory, where they were coupled into single-mode fibres. The fibres then fed the spatially filtered beams into the IONIC3 integrated optics beam combiner \citep{Berger:2003}, which combined the beams coaxially and pairwise with a ratio of 50:50. For each baseline, the beam combination produced two complementary outputs, which were shifted in phase by $\pi$ with respect to each other and which were recorded on a PICNIC camera \citep{Pedretti:2004}. Although the information recorded by these two channels is in principle redundant, it is used to remove residual photometric fluctuations simply by subtracting the signals from the two channels. The PICNIC detector array is a 256 $\times$ 256 pixel array, arranged in four quadrants of 128 $\times$ 128 pixels and sensitive to the 0.8-2.5 $\mu$m wavelength region. The pixel readouts were performed in a non-destructive way.

\begin{figure}[!t]
\begin{center}
\includegraphics[width=8.0 cm]{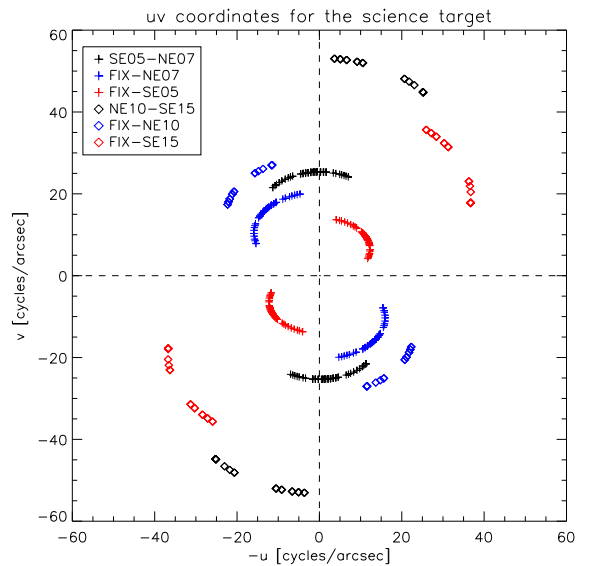}
\caption{Sampling of the Fourier (u,v) plane for the selected data set.}
\label{fig:uv-plane}
\end{center}
\end{figure}

Interferometric observations of Vega were obtained in the infrared H-band (1.50 - 1.80\,$\mu$m) with IONIC, using the SE05-NE07-FIX and NE10-SE15-FIX triplets of the IOTA array (see Figure~\ref{fig:IOTA} and the corresponding u-v plane in Figure~\ref{fig:uv-plane}). Observations took place in June 2006, on June 9, June 10 (NE07-SE05-FIX triplet), June 13 and June 14 (NE10-SE15-FIX triplet). A fringe-tracking algorithm was applied in real time to ensure that interference occurs nearly simultaneously for all baselines \citep{Pedretti:2005}. In addition, the light was split into two orthogonal polarisation axes with a Wollaston prism after beam combination, and these were recorded simultaneously. This choice was initially made to improve the stability of the instrumental response, hence the data quality. However, a non-linear regime of the detector, probably related to partial saturation effects, was observed for several pixels in the data from June 9 and June 10. This non-linearity arose only on the pixels corresponding to the first polarisation, which presents significantly higher fluxes than the second one. Consequently, the first polarisation in the data from June 9 and June 10 has been discarded. This behaviour has not been observed in the data of June 13 and June 14 in any of the two polarisations.

\subsection{Data reduction}

\begin{figure}[!t]
\centering
\includegraphics[width=9 cm]{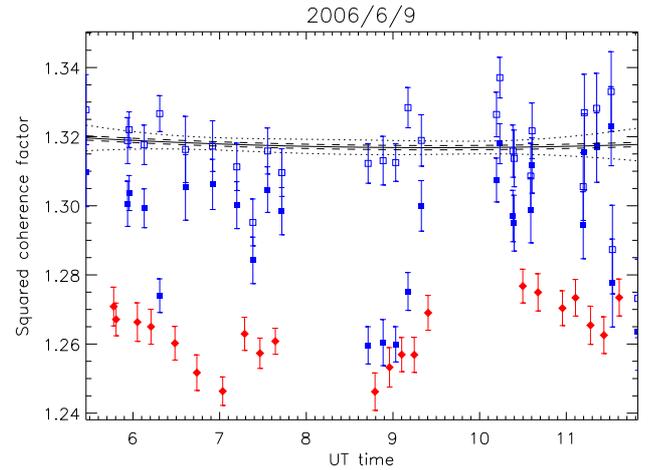}
\caption{Squared coherence factors ($\mu^2$) and interferometric transfer function (TF) estimations (T$^2$) for one representative observing night (June 9th) on the NE07-SE05 baseline. Filled symbols correspond to $\mu^2$ data points for Vega (red diamonds) and the calibrator targets (blue squares). The interferometric TF is represented by the empty blue squares, which are derived from the calibrator measurements taking their known diameter into account. The solid line is an interpolation of the TF, obtained by fitting a second-degree polynomial to its estimated values. The 1-$\sigma$ statistical and systematic error bars on the estimation of the TF are, respectively, represented by dotted and dashed lines. The two gaps during the night (one around 8:00 UT and the other around 9:30 UT) are due to the change in the long delay line offset and to some passing clouds.}
\label{fig:TF}
\end{figure}

The raw squared visibilities and closure phases were obtained by using the established IDL routines described by \cite{Monnier:2004,Monnier:2006}. The interferometric transfer function of the instrument, i.e., the response of the system to a point source, is estimated with calibrator stars. All calibrator stars (listed in Table~\ref{tab:calib}) were chosen from two catalogues developed for this specific purpose \citep{Borde:2002,Merand:2005}. The statistical and systematic error bars on the transfer function estimation were computed following Appendix C of \cite{Kervella:2004c}. Based on the calibrator measurements, the interferometric transfer function was estimated for the whole night by polynomial interpolation \citep[as in][]{Lebouquin:2006, Absil:2009}.  The raw squared coherence factors $\mu^2$ for Vega and its calibrators are shown in Figure~\ref{fig:TF} for a representative night. A second-order polynomial was found appropriate to following the variations in the interferometric transfer function during the whole night as the target and calibrator moved across the sky. The statistical error bar on the interferometric transfer function was computed locally at any given time by using the covariance matrix on the parameters of the polynomial fit. The systematic error bar, on the other hand, was computed globally with a weighted sum of the systematic error bars on all transfer function estimations, taking the correlation between calibrators into account.

\begin{table*}[hbpt]
\begin{center}
\caption{Individual measurements obtained with IOTA/IONIC in June 2006. Only one polarisation is kept in the data of June 9 and 10, while both polarisations are kept for June 13 and 14.}\label{tab:data_overview}
       \begin{tabular}{c r r c c c r r r c c }
       \hline
       \hline
     &  UT     &     HA      & \multicolumn{2}{c}{\# files} &   Array     & \multicolumn{3}{c}{\,\,Proj. baseline [m]} &               &         \\
Date & [hh:mm] &  [hh:mm]    & rec. & pre. & configuration &   AB     &    BC               &  AC   & CAL$_{\rm 1}$ & CAL$_{\rm 2}$ \\
       \hline
2006/06/09&5:47&-3:04&2&2&NE07 SE05 FIX&8.22&4.83&5.86&$\theta$ Her&$\theta$ Her\\
					&6:03&-2:48&1&1&NE07 SE05 FIX&8.29&4.86&6.04&$\theta$ Her&$\theta$ Her\\
					&6:12&-2:38&1&1&NE07 SE05 FIX&8.32&4.87&6.14&$\theta$ Her&$\pi$ Her\\
					&6:29&-2:22&1&1&NE07 SE05 FIX&8.37&4.89&6.30&$\pi$ Her&$\lambda$ Lyr\\
					&6:44&-2:07&1&1&NE07 SE05 FIX&8.41&4.92&6.43&$\lambda$ Lyr&$\lambda$ Lyr\\
					&7:02&-1:49&1&1&NE07 SE05 FIX&8.45&4.94&6.57&$\lambda$ Lyr&$\lambda$ Lyr\\
					&7:17&-1:34&1&1&NE07 SE05 FIX&8.48&4.96&6.67&$\lambda$ Lyr&$\lambda$ Lyr\\
					&7:28&-1:23&1&1&NE07 SE05 FIX&8.50&4.98&6.73&$\lambda$ Lyr&$\lambda$ Lyr\\
				  &7:39&-1:12&1&1&NE07 SE05 FIX&8.51&4.99&6.79&$\lambda$ Lyr&$\lambda$ Lyr\\
					&8:48&0:03&1&1&NE07 SE05 FIX&8.56&4.99&7.01&$\pi$ Her&$\pi$ Her\\
					&8:58&0:07&1&1&NE07 SE05 FIX&8.57&4.98&7.02&$\pi$ Her&$\pi$ Her\\
					&9:06&0:16&1&1&NE07 SE05 FIX&8.57&4.97&7.03&$\pi$ Her&$\pi$ Her\\
					&9:15&0:25&1&1&NE07 SE05 FIX&8.57&4.95&7.04&$\pi$ Her&$\theta$ Her\\
					&9:24&0:34&1&1&NE07 SE05 FIX&8.57&4.93&7.04&$\theta$ Her&$\theta$ Her\\
					&10:30&1:40&1&1&NE07 SE05 FIX&8.56&4.68&7.01&$\theta$ Her&$\kappa$ Lyr\\
					&10:40&1:50&1&1&NE07 SE05 FIX&8.55&4.62&7.00&$\kappa$ Lyr&$\lambda$ Lyr\\
					&10:57&2:07&1&1&NE07 SE05 FIX&8.54&4.51&6.98&$\kappa$ Lyr&$\lambda$ Lyr\\
					&11:06&2:16&1&1&NE07 SE05 FIX&8.53&4.45&6.97&$\kappa$ Lyr&$\lambda$ Lyr\\
					&11:17&2:27&1&1&NE07 SE05 FIX&8.52&4.37&6.96&$\lambda$ Lyr&$\lambda$ Lyr\\
					&11:26&2:36&1&1&NE07 SE05 FIX&8.51&4.30&6.95&$\lambda$ Lyr&$\kappa$ Lyr\\
					&11:37&2:47&1&1&NE07 SE05 FIX&8.50&4.21&6.94&$\kappa$ Lyr&$\kappa$ Lyr\\
2006/06/10&8:10&0:37&3&2&NE07 SE05 FIX&8.55&5.01&6.93&$\theta$ Her&$\theta$ Her\\
					&8:20&0:27&3&3&NE07 SE05 FIX&8.55&5.00&6.96&$\theta$ Her&$\theta$ Her\\
					&8:31&0:16&3&3&NE07 SE05 FIX&8.56&5.00&6.98&$\theta$ Her&$\theta$ Her\\
					&8:42&0:05&3&3&NE07 SE05 FIX&8.56&4.99&7.00&$\theta$ Her&$\theta$ Her\\
					&8:52&0:06&3&3&NE07 SE05 FIX&8.57&4.98&7.02&$\theta$ Her&$\theta$ Her\\
					&9:06&0:19&3&3&NE07 SE05 FIX&8.57&4.96&7.03&$\theta$ Her&$\theta$ Her\\
					&9:16&0:30&3&2&NE07 SE05 FIX&8.57&4.94&7.04&$\theta$ Her&$\pi$ Her\\
					&9:32&0:46&3&3&NE07 SE05 FIX&8.57&4.90&7.04&$\pi$ Her&$\pi$ Her\\
					&9:45&0:58&3&3&NE07 SE05 FIX&8.57&4.86&7.04&$\pi$ Her&$\pi$ Her\\
					&9:55&1:09&3&3&NE07 SE05 FIX&8.57&4.82&7.03&$\pi$ Her&$\pi$ Her\\
2006/06/13&9:35&1:01&2&2&NE10 SE15 FIX&17.72&14.56&10.01&$\theta$ Her&$\kappa$ Lyr\\
          &9:47&1:13&2&1&NE10 SE15 FIX&17.66&14.42&10.00&$\kappa$ Lyr&$\kappa$ Lyr\\
					&10:01&1:27&2&1&NE10 SE15 FIX&17.58&14.24&9.98&$\kappa$ Lyr&$\theta$ Her\\
					&10:28&1:53&2&2&NE10 SE15 FIX&17.40&13.81&9.95&$\kappa$ Lyr&$\theta$ Her\\
2006/06/14&7:08&-1:23&2&2&NE10 SE15 FIX&17.99&14.92&9.57&$\pi$ Lyr&$\pi$ Her\\
					&7:19&-1:12&2&2&NE10 SE15 FIX&17.98&14.96&9.65&$\pi$ Lyr&$\pi$ Her\\
					&7:31&-0:60&2&2&NE10 SE15 FIX&17.98&14.98&9.73&$\pi$ Lyr&$\theta$ Her\\
					&7:51&-0:40&2&2&NE10 SE15 FIX&17.96&15.01&9.84&$\theta$ Her&$\theta$ Her\\
					&8:01&-0:29&2&2&NE10 SE15 FIX&17.95&15.01&9.88&$\theta$ Her&$\kappa$ Lyr\\
					&10:23&1:52&2&2&NE10 SE15 FIX&17.41&13.83&9.95&$\theta$ Her&$\theta$ Her\\
\hline
\end{tabular}\\
\end{center}
{\small Cols. 1, 2, and 3 give the date, UT time, and the hour angle, cols. 4 and 5 give the corresponding number of files recorded and kept during the data reduction process, col. 6, 7, 8, and 9 give the array configuration and corresponding projected baselines, respectively. The last two columns give the calibrator stars used before (CAL$_{\rm 1}$) and after (CAL$_{\rm 2}$) the observation of Vega.}
\end{table*}

Since all calibrators chosen in this study are late K giants, the squared visibilities and closure phases were computed for the wide spectral bandwidth on which the IOTA observations were performed, considering the actual spectrum of the star \citep[using tabulated H-band spectra from][]{Pickles:1998} and the spectral transmission of the IOTA/IONIC instrument. An additional systematic error was considered in order to consider the chromaticity effects that limit the absolute precision of calibration. For squared visibilities, we estimated the impact of chromatism using spectrally dispersed data that were obtained the same year with IONIC \citep[see][]{Lacour:2008,Pedretti:2009}. Our analysis shows that the transfer function does not vary significantly with the spectral type of the observed star for any given instrumental setup, but that the shape of the transfer function with respect to wavelength depends on the instrumental set-up (baseline, detector, read-out mode, etc.) in an unpredictable way. Therefore, we cannot accurately predict the colour correction factor for V$^2$ measurements between the various observed spectral types. This variable behaviour of the spectral transfer function across the H band has an impact on the calibration accuracy of $\pm$1\% at most between an A0V and a K3III star, which we add quadratically to our estimated error bars. For closure phases, we used a conservative systematic error bar of 0.5$^\circ$ to account for chromatism effects. This value is based on engineering tests carried out by \cite[][]{Monnier:2006}, which showed that the closure phase presents a systematic error of this order between a hot star (B8) and a cool star (M3).

Finally, for the nights of June 13 and 14 (large triplet), only data corresponding to hour angles ranging between -2 and 2\,hours have been used in the following discussions to achieve good accuracy on the calibrated squared visibilities (see individuals measurements in Table~\ref{tab:data_overview}). The final calibrated data are presented in Figure~\ref{fig:vega_reduc} (squared visibility) and Figure~\ref{fig:vega_closure} (closure phase).

\begin{table*}[!t]
\begin{center}
\caption{Fundamental parameters and estimated angular diameter for Vega ($\alpha$ Lyr) and its calibrators.}\label{tab:calib}
    \begin{tabular}{c c c c c c c c c c}
        \hline
        \hline
        Identifier & HD number & RA-J2000 & DEC-J2000 & Sp. type & $m_V$ & $m_H$ & $\theta_{\rm LD}\pm1\sigma$ & Refs. \\
        & & [d m s] & [d m s] & & & & [mas] & \\
        \hline
        $\pi$ Her & 156283 & 17 15 02.83 & +36 48 32.98 & K3II & 3.16 & 0.12 & 5.29$^{0.055}$ & [1,2]\\
        $\theta$ Her & 163770 & 17 56 15.18 & +37 15 01.94 & K1IIa & 3.85 & 1.14 & 3.15$^{0.034}$  & [1,2]\\
        $\kappa$ Lyr & 168775 & 18 19 51.71 & +36 03 52.37 & K2IIIab & 4.32 & 1.86 & 2.28$^{0.025}$ & [1,2]\\
        $\alpha$ Lyr & 172167 & 18 36 56.33 & +38 47 01.29 & A0V & 0.03 & -0.03 & 3.312$^{0.067}$ & [1,3]\\
        $\lambda$ Lyr & 176670 & 19 00 00.83 & +32 08 43.85 & K2.5III & 4.95 & 1.84 & 2.41$^{0.026}$ & [1,2]\\
    \hline
    \end{tabular}\\
\end{center}
    {\small References. Coordinates, spectral types and magnitudes from [1] SIMBAD; limb-darkened diameters and 1-$\sigma$ errors from [2] \cite{Borde:2002}, and [3] \cite{Absil:2007b}.}
\end{table*}

\section{Modelling the data with simple models}

As shown by \cite{Absil:2006b} with CHARA/FLUOR, a realistic stellar photospheric model cannot reproduce K-band visibility measurements. As shown in Figure~\ref{fig:vega_reduc}, this is also the case for our H-band IOTA/IONIC data with measured visibilities clearly below the expected level of the photospheric visibility. These observations cannot be reconciled with a purely photospheric model, because on the one hand short-baseline data are weakly sensitive to the model parameters (angular diameter and limb-darkening profile), and on the other, the model parameters are already known with good accuracy \citep{Aufdenberg:2006,Peterson:2006}. Furthermore, the visibility deficit is observed on the three different baselines with roughly the same magnitude so that it cannot be explained by the apparent oblateness of Vega's photosphere (which is not significant since Vega is seen almost pole-on).

To explain this visibility deficit, \cite{Absil:2006b} proposed a model of a star surrounded by an exozodiacal disc accounting for 1.26 $\pm$ 0.28\% of the K-band stellar flux within the FLUOR field-of-view \citep[$\sim$8\,AU in radius,][]{Absil:2007b}. In particular, they addressed and ruled out a series of scenarios. For instance, a bound low-mass companion was also shown to be a viable explanation although much less probable. Other explanations, such as a bright background object within the FLUOR field-of-view (probability less than 10$^{-6}$) or stellar spots (appearing in the second and higher lobes of the visibility function), were ruled out with good confidence. At that time, it was argued that stellar winds and mass-loss events could also be the origin of the near-infrared excess, but these scenarios were shown to be very unlikely later by \cite{Absil:2007b}. In the two following sections, we consider the two most probable interpretations separately (the exozodiacal disc and the binary companion) and check whether simple models can reproduce our measurements.

\subsection{The exozodiacal disc scenario}\label{sec:disc_interfero}

\begin{figure}[!t]
\centering
\includegraphics[width=9.0 cm]{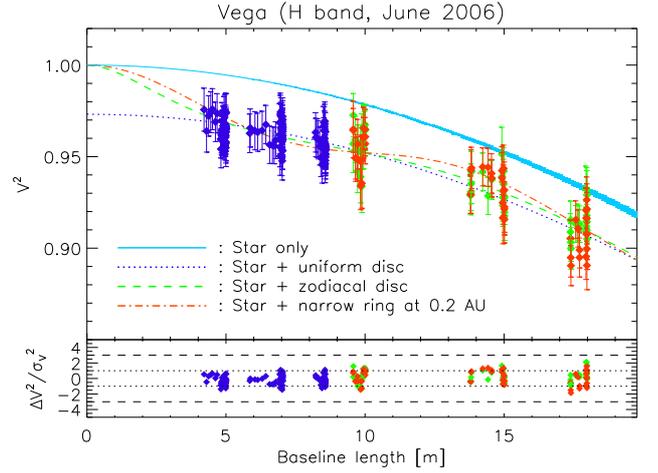},
\caption{Expected squared visibility as a function of the baseline length for Vega's limb darkened photosphere (blue solid line), given with the measured squared visibilities and related error bars (blue symbols for June 9 and 10, green symbols for June 13 and 14 on polarisation 1, and red symbols for June 13 and 14 on polarisation 2). The thickness of the blue solid line corresponds to the 5-$\sigma$ uncertainty related to the uncertainty on the stellar diameter. The best-fitted models of a limb-darkened photosphere surrounded by a uniform circumstellar emission (dotted blue line), a zodiacal disc (dashed green line), and a narrow ring (dash-dotted red line) are also represented (see discussion in Section 3). Residuals of the fit are given in the bottom panel for the uniform disc model.}\label{fig:vega_reduc}
\end{figure}

To constrain the near-infrared excess emission around Vega, three simple models of a circumstellar disc have been fitted to the data: a diffuse emission uniformly distributed across the field-of-view, the zodiacal disc model of \cite{Kelsall:1998}, which is implemented in the Zodipic package,\footnote{Zodipic is an IDL program developed by M.~Kuchner and C.~Stark for synthesizing images of exozodiacal clouds. It can be downloaded from http://asd.gsfc.nasa.gov/Marc.Kuchner/home.html.} and a narrow ring of dust located at twice the sublimation radius ($r_{\rm sub}$=0.1\,AU for a sublimation temperature T$_{\rm sub}$=1700\,K). All models are assumed to be point-symmetric, as suggested by the calibrated closure phases (see Figure~\ref{fig:vega_closure}), and use an effective temperature of 8027\,K for Vega as viewed by the dust in the equatorial plane (Jason Aufdenberg, private communication). The results of this fitting procedure are shown in Figure~\ref{fig:vega_reduc}, with the corresponding H-band disc/star flux ratios and reduced $\chi^2$ in Table~\ref{tab:flux_ratio}. All three models fit the data set ($\chi_r^2\cong$0.5-0.7) equally well with a best-fit flux ratio of 1.35 $\pm$ 0.49\%, 1.26 $\pm$ 0.45\%, and 1.20 $\pm$ 0.43\%, respectively, for the uniform disc model, the zodiacal disc model, and the ring model. All the best-fit contrasts are compatible within 1\,$\sigma$, which indicates that the final flux ratio does not heavily depend on the distribution of the circumstellar emission.

To confirm this conclusion, the squared visibilities from synthetic images of a geometrically and optically thin debris disc were computed and compared to the data set. The synthetic images are based on three parameters: the disc/star flux ratio, the inner disc radius $r_{\rm in}$, and the exponent $\alpha$ of the power law describing the density decrease as a function of distance ($n_r\propto r^{\alpha}$). The synthetic images only include thermal emission, which is expected to be largely dominant for the dust temperatures explored here, and they assume pure blackbody emission (grain temperature proportional to $r^{-0.5}$, with $T$ = 1700\,K at 0.1\,AU). For each couple of parameters ($r_{\rm in}$,$\alpha$) ranging between the sublimation radius and 2\,AU and between $0$ and $-50$,  respectively, the best-fit flux ratio was computed by $\chi^2$ optimisation. The results show that all models fit the data equally well with a reduced $\chi^2$ ranging between 0.48 and 0.84, confirming exozodiacal dust as a good scenario to explain the deficit of visibility. It is, however, impossible to conclude on the location of the dust based on the sole interferometric data. The best-fit disc/star flux ratios are all compatible within approximately 1-$\sigma$ error bars and range from 1.04\% to 1.42\%. We therefore adopt a H-band flux ratio of 1.23 $\pm$ 0.53\% in the following discussions. The error bar includes both the uncertainty due to the model ($\pm$0.19\%) and the maximum error bar on an individual model ($\pm$0.49\%). This inferred disc/star contrast is marginally compatible with the one predicted by \cite{Absil:2006b} based on the CHARA/FLUOR observations (about 0.6\% with an inner rim located between 0.17 and 0.3\,AU). The implications on the disc properties of this new H-band measurement are discussed in Section~\ref{sec:modeling}.

\begin{table}[!t]
\begin{center}
\caption{Best-fit flux ratio and goodness of fit for 3 different disc models.}\label{tab:flux_ratio}
		\begin{tabular} {l c c}
		\hline
		\hline
		Disc model & Flux ratio [\%] & $\chi^2_r$ \\
		\hline
		Uniform emission & 1.35 $\pm$ 0.49 & 0.52 \\
		Zodiacal disc & 1.26 $\pm$ 0.45 & 0.52 \\
		Narrow ring [0.2\,AU] & 1.20 $\pm$ 0.43 & 0.62 \\
		\hline	
		\end{tabular}
\end{center}
{\small The angular diameter and effective temperature of the stellar photosphere have been fixed respectively to 3.312 $\pm$ 0.067\,mas (see Table~\ref{tab:calib}) and 8027\,K as viewed by the dust in the equatorial plane (see main text).}
\end{table}

\subsection{The binary scenario}

The possible presence of a low-mass binary companion around Vega has been the subject of various studies (e.g., radial velocities, astrometric measurements, and high dynamic range single-pupil imaging),  drastically reducing the range of allowed parameters for the putative companion.  Before discussing the existing data in the literature, it is necessary to check whether a binary companion could effectively reproduce our interferometric observations. Given the relatively good sampling of the u-v plane (see Figure~\ref{fig:uv-plane}) and the uniform deficit measured along the different baselines (see Figure~\ref{fig:vega_reduc}), we might expect to derive strong constraints on the properties of a binary companion based on the sole interferometric measurements.

\begin{figure}[!t]
\centering
\includegraphics[width=9.0 cm]{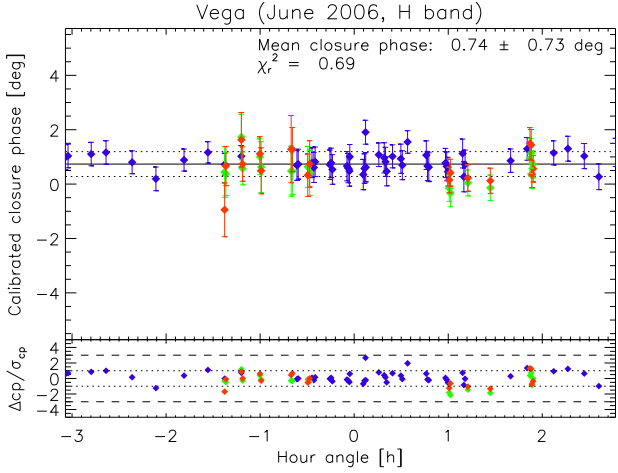}
\caption{Calibrated closure phase measurements given with error bars as a function of hour angle. The color code is the same as in Figure~\ref{fig:vega_reduc}. The mean closure phase is represented by the black solid line, while the dotted lines represent the 1-$\sigma$ statistical uncertainty on its value. The bottom plot shows the residual between the data and the mean value of closure phases.}
\label{fig:vega_closure}
\end{figure}

\begin{figure*}[!t]
\centering
\includegraphics[width=8.0 cm]{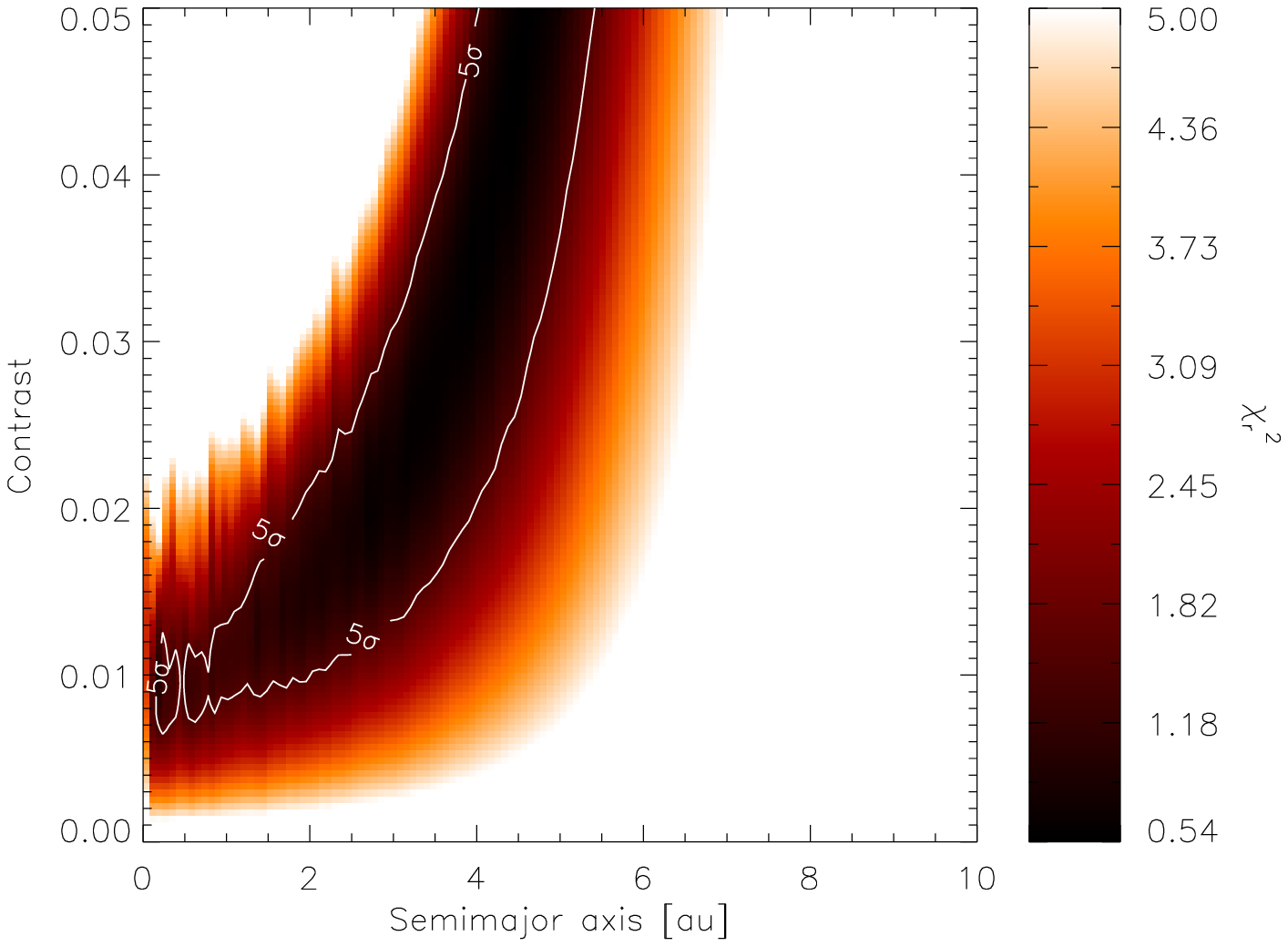}
\includegraphics[width=8.0 cm]{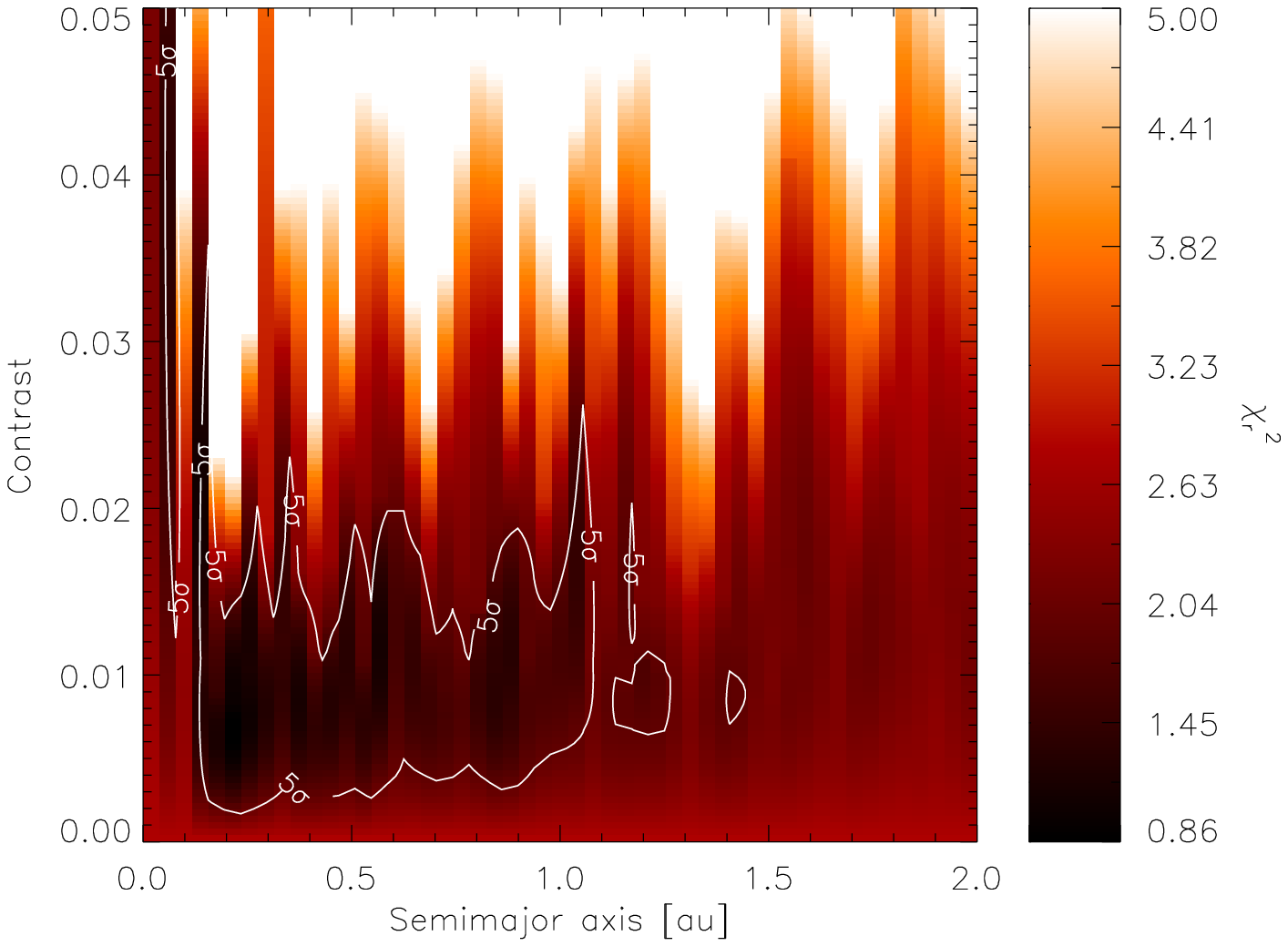}
\caption{Reduced $\chi^2$ map for the fit of a simple binary model to the squared visibility (left) and closure phase (right). The maps are given as a function of two free parameters: the semi-major axis of the companion's orbit and the flux ratio between the putative companion and Vega. For each couple of semi-major axis/contrast, the orbital phase at a given time t$_0$ has been optimised to minimize the $\chi_r^2$. All models corresponding to a semi-major axis larger than 2\,AU in the closure phase map are beyond 5\,$\sigma$ and are not shown for the sake of clarity.}\label{fig:cp_maps}
\end{figure*}

To assess whether a low-mass binary companion would be compatible with our observations, we focus as a first step on the calibrated closure phases, which are shown in Figure~\ref{fig:vega_closure} as a function of the datafile number. The mean closure phase is 0.74$^\circ$, with a 1-$\sigma$ error of $\pm$0.73$^\circ$. This error includes the statistical error due to the scatter in the individual CP measurements and the systematic error due to the chromatism of the beam combiner (0.5$^\circ$ as discussed previously), which clearly contributes to the low positive value of the mean closure phase. Basically all closure phases are consistent with the mean value ($\chi_r^2$ of 0.69) for all the various projected baselines of both configurations, suggesting that any deviation from point symmetry in the
near-infrared excess source is not larger than 1\% for the spatial frequencies explored here. To determine which combination of orbital parameters and contrast of the binary companion would be compatible with these measurements, we fitted a model of a binary star to the whole data set (squared visibilities and closure phases) in a second step. The orbital motion of the putative stellar companion is taken into account assuming a mass of 2.3\,$M_\odot$ for Vega \citep{Aufdenberg:2006}. The mass of the companion is computed using the evolutionary models developed by \cite{Baraffe:1998}, assuming it has the same age as Vega itself \citep[about 455\,Myr,][]{Yoon:2010}. We also assumed the companion to be orbiting on a circular orbit within the plane of the sky \citep[Vega is almost seen pole-on with an inclination angle estimated to 5.7$^\circ$,][]{Aufdenberg:2006}. The model is therefore based on three parameters: the semi-major axis, the orbital phase at a given time t$_0$, and the binary flux ratio. For each couple of semi-major axis/contrast, we computed the orbital phase at t$_0$ that minimizes the $\chi_r^2$. With this, we have produced $\chi_r^2$ maps, which are represented in Figure~\ref{fig:cp_maps} as a function of semi-major axis and binary contrast for the squared visibilites (left) and the closure phases (right).

The $\chi_r^2$ maps show that a range of models fits the interferometric data in a satisfactory way, the effect of the Gaussian beam profile being clearly evident in the lefthand figure as the binary contrast required to fit the data increases with the binary separation. All models corresponding to a binary separation larger than 1.5\,AU can be rejected with a 5-$\sigma$ confidence level by the closure phase map. Within 1.5\,AU, the squared visibility map shows that only models corresponding to a contrast between 0.6 and 1.6\% are compatible with the data with the same confidence level. Furthermore, a very short semi-major axis ($<$0.1\,AU) can be rejected because it is either incompatible with the squared visibilities or with the closure phases, regardless of the contrast. Therefore, based on our interferometric data, the only possible parameters for a putative binary companion are a semi-major axis ranging between 0.1 and 1.5\,AU and a contrast ranging between 0.6 and 1.6\%. Such properties for the companion are, however, rejected by Hipparcos astrometric measurements of Vega, which have shown that the orbital semi-major axis of a putative companion cannot be larger than 6.3\,mas (=0.05\,AU=4\,R$_\star$) with a 99\% confidence assuming a circular orbit \citep{Perryman:1997,Absil:2006b}. The absence of a detected companion in radial velocity measurements confirms this conclusion further. All these elements suggest that the presence of a close companion around Vega to explain the observed visibility deficit is extremely unlikely.

\section{Further constraints on the exozodiacal disc}\label{sec:modeling}

The analysis presented in the previous section confirms that the most straightforward scenario for explaining the visibility deficit is the presence of hot dust grains producing thermal emission in the near-infrared. To constrain the disc parameters, several archival spectro-photometric and interferometric measurements at near- and mid-infrared wavelengths exist in the literature (see Table~\ref{tab:fluxes}). In addition to the measurements used by \cite{Absil:2006b}, we add our direct measurement of the H-band excess flux (1.23 $\pm$ 0.53\%) and the IRS spectrum found in the Spitzer archives. The Spitzer spectrum has been reduced with the c2d pipeline developed by \cite{Lahuis:2006}, and binned into a few equivalent broad-band photometric measurements for the sake of spectral energy distribution (SED) modelling. The idea is then to check whether there is at least a disc model that is compatible with all the constraints and derive its properties.

To reproduce the SED of the infrared excess as listed in Table~\ref{tab:fluxes}, radiative transfer modelling was performed using the code developed by \cite{Augereau:1999} for cold debris discs, and adapted to the case of exozodiacal discs \citep[e.g.,][]{Absil:2006b,Difolco:2007,Absil:2007b,Absil:2009}. The code considers a population of dust grains with a parametric surface density profile and size distribution, and computes the thermal equilibrium temperature of the grains exposed to the stellar radiation. It can handle various chemical compositions for the grains in order to investigate for instance the fraction of carbonaceous or silicate material contributing to the observed emission. Specific care is given in the model to the treatment of exozodiacal dust close to the sublimation radius, to account for the size-dependent position of this radius. On output, the code computes the thermal and scattered light emissions of debris disc models and produces both images and SEDs over a broad parameter space to combine observations from different instruments (IOTA/IONIC, CHARA/FLUOR, MMT/BLINC, Spitzer/IRS, broad-band fluxes), taking their specific response into account.

\begin{figure}[!t]
\centering
\includegraphics[width=9.0 cm]{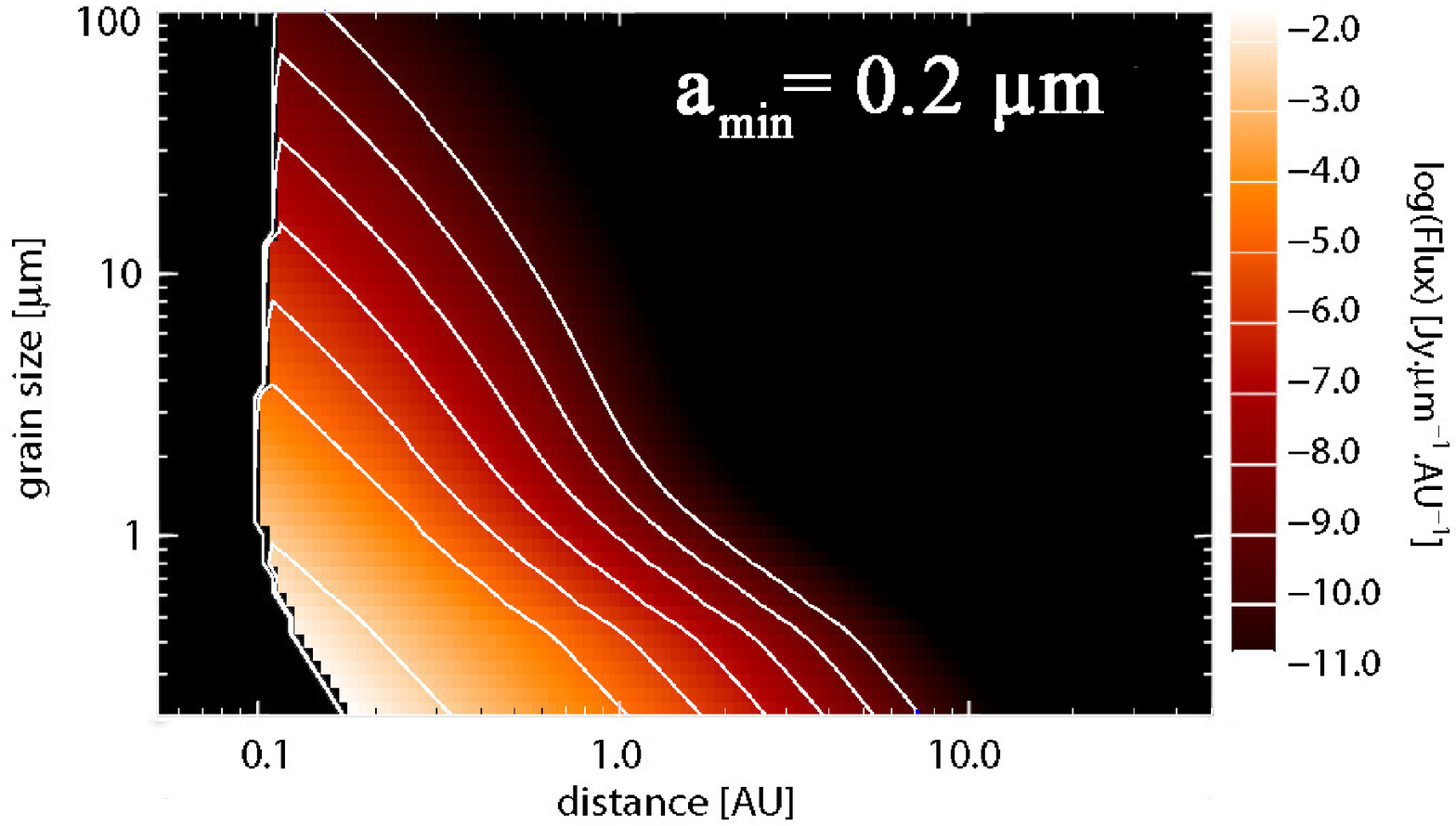}
\includegraphics[width=9.0 cm]{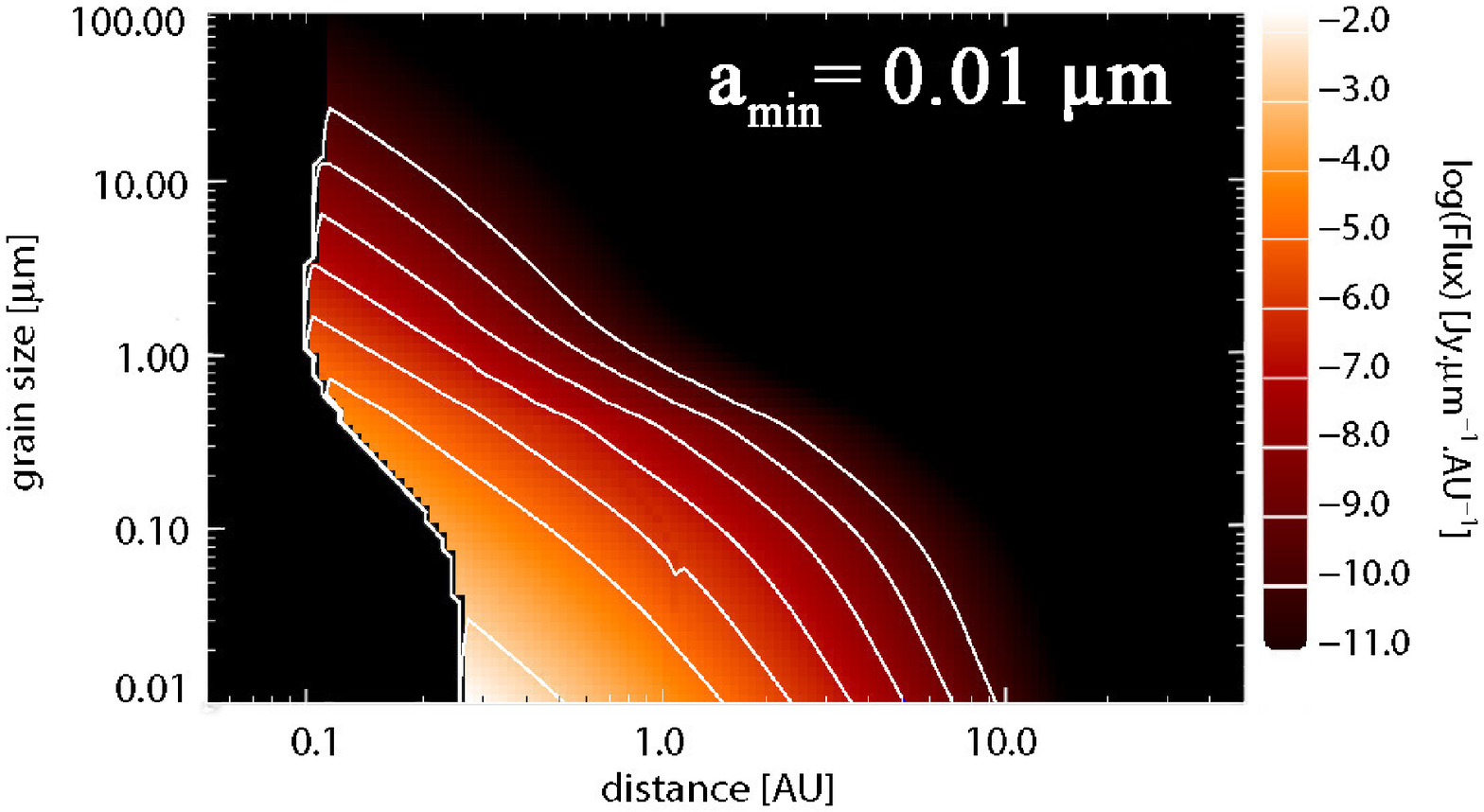}
\caption{Flux density maps given as a function of grain size and distance to the star computed for $a_{\rm min}$ = 0.2\,$\mu$m (best-fit model, upper figure) and $a_{\rm min}$ = 0.01\,$\mu$m  (lower figure). A surface density power law of $\alpha=-3$ and a size distribution power-law exponent of $\kappa=-5$ have been assumed in these plots, with a maximum size of 1000\,$\mu$m. In this simulation, the disc is composed of 50\% silicates and 50\% carbonaceous grains ($C_r$=50\%, see text).  The contours are plotted every power of 10 between 10$^{-1}$ and 10$^{-7}$ of the maximum flux density.} \label{fig:chi2_model}
\end{figure}

Using the code described above with an effective temperature for Vega of 8027\,K as viewed by the dust in the equatorial plane, a grid of debris disc models was computed with the following parameters: the peak position $r_0$ of the surface density profile (possible values ranging from 0.05 to 1\,AU, provided that the sublimation radius is not reached), the minimum size $a_{\rm min}$ of the size distribution (from 0.01 to 54.6\,$\mu$m), the slope $\kappa$ of the size distribution (from -2.7 to -9), the slope $\alpha$ of the surface density profile beyond $r_0$ (from 0 to -9.5), and the volume fraction $C_r$ of carbon grains (from 0 to 100\%). Each model assumes no azimuthal dependence and sublimation temperatures of 1200\,K for silicate grains \citep[``astrosilicates'',][]{Draine:2003} and 1900\,K for carbonaceous grains \citep[``amorphous carbon'',][]{Zubko:1996}. The size distribution (between $a_{\rm min}$ and $a_{\rm max}$) has been accordingly truncated at the sublimation size ($a_{\rm sub}$), which depends on the radial distance to the star since only grains large enough to survive the sublimation process can actually coexist (see Figure~\ref{fig:chi2_model}).

\begin{table}[!t]
	  \caption{Available constraints on the near- and mid-infrared excess around Vega.}\label{tab:fluxes}
	  \renewcommand{\tabcolsep}{3pt}
		\begin{center}
		\begin{tabular}{c c c c}		
		\hline
		\hline
		Wavelength & F$_{\rm meas}$ & Excess & Instruments\\
		$[\mu m]$ & [Jy] & [\%] & \\
		\hline
		1.26 & 1574 $\pm$ 34 & 2.4 $\pm$ 2.9\% & Catalina$^1$, UKIRT$^2$\\
		1.60 & 1055 $\pm$ 32 & -2.4 $\pm$ 3.6\% &  Catalina$^1$ \\
	  1.65 & -- & 1.23 $\pm$ 0.53\% & IOTA/IONIC \\
		2.12 & -- & 1.29 $\pm$ 0.19\% & CHARA/FLUOR$^3$\\
		2.12 & -- & $5^{+1}_{-2}\%$ & PTI$^4$\\
		2.20 & 655 $\pm$ 14 & 4.1 $\pm$ 3.0\% & Catalina$^1$, UKIRT$^2$\\
		3.54 & 283 $\pm$ 6 & 3.1 $\pm$ 3.0\% & Catalina$^1$, UKIRT$^2$\\
		4.80 & 167 $\pm$ 8 & 7.1 $\pm$ 5.1\% & Catalina$^1$, UKIRT$^2$\\
		10 & 40.0 $\pm$ 3 & 6.0 $\pm$ 4.5\% & Various$^5$\\
		10.5 & 33.7 $\pm$ 0.33 & 0.0 $\pm$ 5.0\% & IRS spectrum\\
		10.6 & -- & 0.2 $\pm$ 0.7\% & MMT/BLINC$^6$\\
		12 & 27.0 $\pm$ 0.3 & 1.2 $\pm$ 5.0\% & IRAS$^7$\\
	  12.5 & 23.89 $\pm$ 0.39 & 0.0 $\pm$ 5.0\% & IRS spectrum\\
		\hline	
		\end{tabular}
		\end{center}
{\small The superscript on the name of the instrument in the last column gives the reference: (1) \cite{Campins:1985}, (2) \cite{Blackwell:1983}, (3) \cite{Absil:2007b}, (4) \cite{Ciardi:2001}, (5) \cite{Rieke:1985}, (6) \cite{Liu:2004}, (7) \cite{Cohen:1992}, with the absolute photometric error estimated by \cite{Aumann:1984}. The interferometric data from IONIC, FLUOR, PTI, and BLINC only sample a specific part of the inner disc, while the photometric studies include Vega's entire environment.}
\end{table}

\begin{figure}[!t]
\centering
\includegraphics[width=8.0 cm]{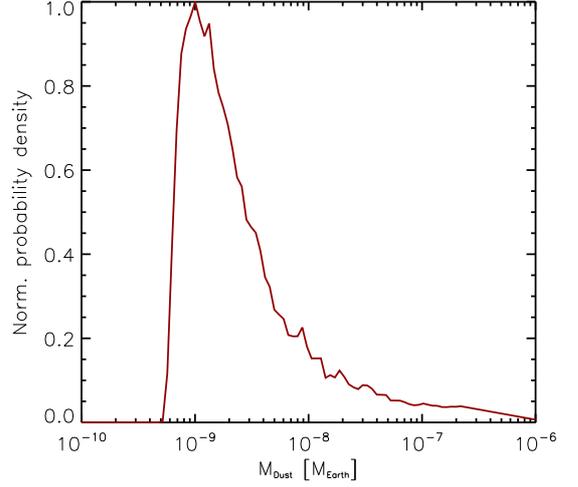}
\caption{Normalised probability density for the disc total mass computed with a Bayesian approach based on the $\chi^2$ grid. The probability density peaks at about $10^{-9}M_\oplus$, equivalent to the mass of an asteroid about 20\,km in diameter.} \label{fig:bayes_mass}
\end{figure}

For each model on the grid, the SED has been computed and compared to the measurements reported in Table~\ref{tab:fluxes}. The first obvious result obtained from the $\chi_r^2$ analysis is that pure silicate models ($C_r$=0) do not fit the SED well with a $\chi_r^2$ always larger than 7. The goodness of the fit improves significantly by introducing carbonaceous grains, which can survive closer to Vega due to their higher sublimation temperature. This conclusion is supported by previous studies explaining the lack of significant silicate emission features around 10\,$\mu$m by the presence of large amounts of highly refractive grains in the inner disc \citep{Laor:1993, Zubko:1996, Gaidos:2004}. The actual amount of carbonaceous grains can, however, not be constrained since the $\chi_r^2$ typically varies from 1.3 for a carbonaceous-poor disc ($C_r$=5\%) to 0.8 for a pure carbonaceous disc ($C_r$=100\%) in the steep power law regime ($\alpha <-3$). For a flatter power law ($\alpha >-3$) like in our zodiacal disc \citep[$\alpha=-0.34$,][]{Kelsall:1998}, the disc models do not fit the various measurements well with a $\chi_r^2$ always higher than 5 even for pure carbonaceous discs. In addition, relatively small grains ($a_{\rm min} <$ 1\,$\mu$m) with a steep size distribution profile ($\kappa<$-3) are also required to obtain a good fit to the SED ($\chi_r^2 <3$). Two good-fit models are represented in Figure~\ref{fig:chi2_model} by flux density maps given as a function of grain size and distance to the star. The upper figure corresponds to the best-fit model ($\chi_r^2=0.84$, total mass = $1.9\times10^{-9} M_\oplus$) obtained for $a_{\rm min}$ = 0.2\,$\mu$m and $\kappa=-5$, while the bottom figure corresponds to the same model with $a_{\rm min}$ equal to 0.01\,$\mu$m ($\chi_r^2=1.01$, total mass = $2.1\times10^{-9} M_\oplus$). The dependence of sublimation temperature on the grain size is clearly visible in these figures, the larger grains surviving closer to Vega than the small ones. The figures also show that the largest grains located near 0.1\,AU contribute relatively less to the total flux than the smaller grains located further away from the star and present in larger numbers. In both cases, the flux density decreases very quickly with the distance to Vega as shown by the various contours. This geometry is mostly constrained by the interferometric measurements, and particularly the non-detection reported by \cite{Liu:2004} with the MMT/BLINC, which requires a steep power law in order to reduce the amount of dust in the regions further than 1\,AU. In addition, recent observations of Vega with the Palomar Fiber Nuller \citep[$\sim$2.2\,$\mu$m,][]{Mennesson:2011} suggest that any dust population contributing to at least 1\% of the near-infrared excess can arise only within 0.2\,AU, a conclusion in favour of $a_{\rm min}$ equal to 0.2\,$\mu$m in order to sufficiently reduce the dust emission beyond 0.2\,AU. Finally, we used a Bayesian approach to compute the normalised probability density of the disc models based on the $\chi^2$ grid. The results of this analysis shows that the disc mass has a probability density that peaks at about $10^{-9}M_\oplus$ (see Figure~\ref{fig:bayes_mass}), equivalent to the mass of an asteroid about 20\,km in diameter (assuming 2.5 g/cm$^3$ density).

\begin{figure*}[!t]
\centering
\includegraphics[height=7cm,width=18.0 cm]{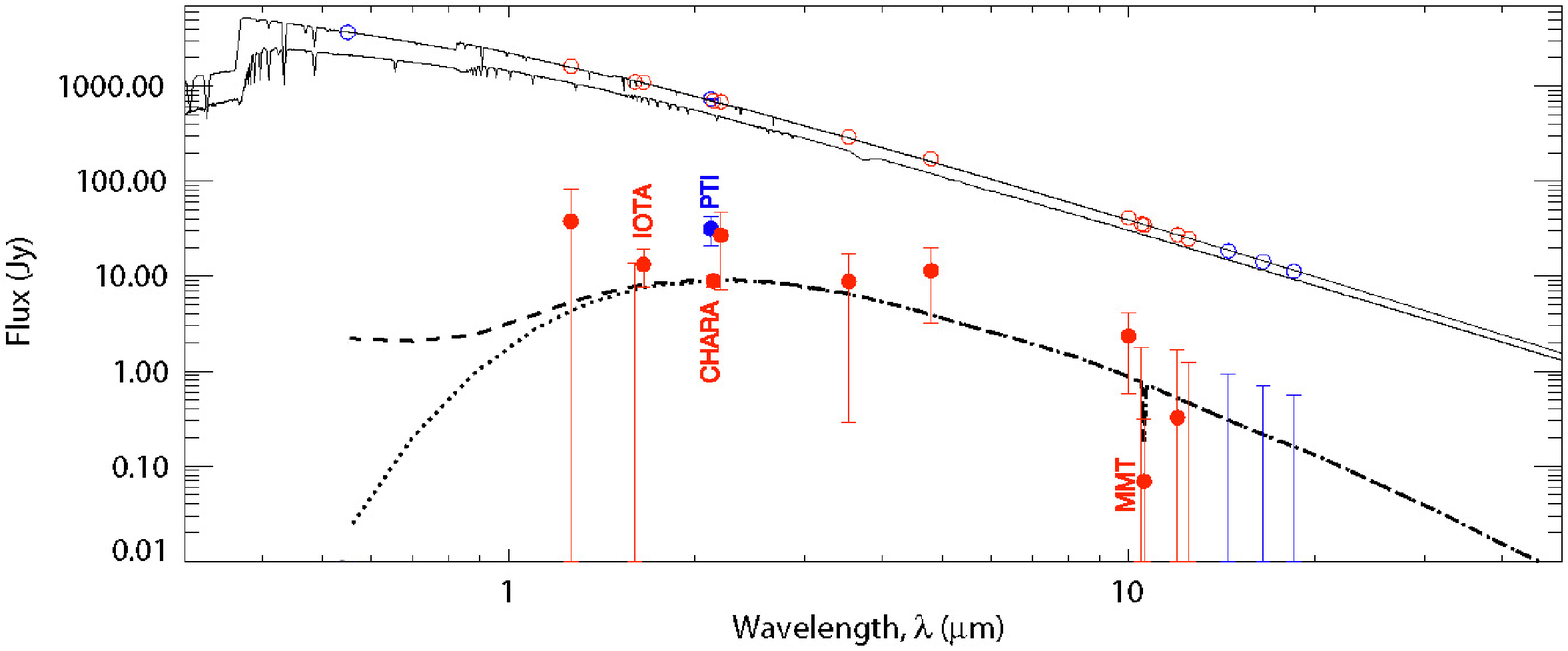}
\caption{A possible good fit of our debris disc model to the photometric and interferometric constraints of Table~\ref{tab:fluxes} ($\chi_r^2 = 0.84$). The model used here has a size distribution exponent of $-5$ with limiting grain sizes of a$_{\rm min}$ = 0.2\,$\mu$m and a$_{\rm max}$ = 1000\,$\mu$m, a surface density power law of $-3$, and it assumes a disc composed of 50\% silicates and 50\% carbonaceous grains. The dashed and dotted lines represent the total emission from the disc on a 6\,AU field-of-view, respectively with and without the scattered emission. The photospheric SED, simulated by a NextGen model atmosphere, is represented by the two solid lines as seen pole-on (upper curve) and from the equatorial plane (lower curve). Only measurements plotted in red have been taken into account in the fit.}\label{fig:vega_model}
\end{figure*}

In summary, the most straightforward scenario consists in a compact debris disc with a steep density profile ($\alpha \le-3$) producing a thermal emission that is largely dominated by small ($a_{\rm min}<1.0$\,$\mu$m) silicates and carbonaceous grains located between 0.1 and 0.3\,AU from Vega. A representative SED is shown in Figure~\ref{fig:vega_model}, together with the SED of the stellar photosphere as seen pole-on (upper solid curve, T$_{\rm eff}$=10150\,K) and as seen by the exozodiacal dust (lower solid curve, T$_{\rm eff}$=8027\,K). This model assumes a size distribution exponent of $-5$ with limiting grain sizes of a$_{\rm min}$ = 0.2\,$\mu$m and a$_{\rm max}$ = 1000\,$\mu$m, a surface density power law of $-3$, and a disc composed of 50\% silicates and 50\% carbonaceous grains. By using this disc model to fit our interferometric data set, we derived a best-fit contrast of 1.23 $\pm$ 0.45\%. Finally, the flux ratio derived here is about 1.5\,$\sigma$ above the value extrapolated from the K-band CHARA/FLUOR measurements \citep[expected H-band excess of 0.6\%,][]{Absil:2006b}. Although this model successfully reproduces the Vega SED, including both CHARA/FLUOR and IOTA/IONIC flux ratios, another possible scenario that could explain the small discrepancy (at least partially) is that the brightness of the exozodiacal disc around Vega is evolving on time scales as short as one year. This issue is discussed briefly in the next section and is currently the subject of follow-up oservations of Vega at the CHARA array.

\section{Discussion}

Several scenarios can explain the presence of hot dust in the close vicinity of Vega. The age of Vega \citep[about 455\,Myr,][]{Yoon:2010} precludes the possibility for this dust to be primordial, because the time scale to remove such dust is shorter than 10\,Myr \citep{Backman:1993,Wyatt:2008}. The detected dust is therefore second generation and produced either by the collision of larger bodies or by cometary sublimation. In addition, a large dust production rate ($\sim10^{-9}M_\oplus$/yr) is necessary to account for the amount of detected dust, which has a limited lifetime (of a few years at most) in the inner Vega's system due to radiation pressure and collisional destruction \citep{Krivov:2000}. As pointed out by \cite{Absil:2006b}, the inward transport of dust from the outer disc \citep[located at about 85\,AU, e.g.,][]{Sibthorpe:2010} through P-R drag is very inefficient given the long time scale of this process \citep[2$\times10^7$\,yr,][]{Dent:2000} and the relatively short collisional time scale in the outer disc (5$\times10^5$\,yr). Instead, the dust would be generated locally by comets orbiting in the inner system. The comets would likely originate from the outer disc of Vega or possibly from  an inner population of icy bodies as in the case of $\beta$ Pic \citep{Beust:2000}. In fact, comets are believed to be the origin of at least 90\% of the dust observed in the solar zodiacal cloud \citep[][]{Nesvorny:2010}. Because they are exposed to the stellar radiation and wind, the dust particles released in the inner Vega system could then pile up in a zone close the sublimation radius with a very steep density profile \citep{Kobayashi:2009}.

In any case, the high production rate needed to account for the amount of detected dust suggests that Vega is currently undergoing major dynamical perturbations. A dynamical ``shake-up'' of the whole planetary system, similar to the late heavy bombardment (LHB) that occurred in the early history of the solar system \citep{Gomes:2005}, could enhance this transfer rate and induce the necessary amount of dust in the inner planetary system \citep[the solar zodiacal cloud is supposed to have been up to $10^4$ times brighter during the LHB,][]{Nesvorny:2010}. Although the presence of giant planets around Vega has not been confirmed yet, such a bombardment would most probably be triggered by the outward migration of giant planets. In particular, \cite{Wyatt:2003} and \cite{Reche:2008} suggest that the outward migration of a Neptune to Saturn-mass body from 40 to 65\,AU could explain the observed clumpy structure in Vega's outer disc. Further dynamical analyses show that such a transport would require the presence of at least two planets, such as a Saturn-like planet orbiting between 40 and 65\,AU and a Jupiter-like planet orbiting closer to Vega (Vandeportal et al. 2011, in preparation).

\section{Conclusion}

Interpreted as the signature of hot exozodiacal dust, the near-infrared excess source detected around Vega by CHARA/FLUOR in the K band is confirmed by our IOTA/IONIC measurements at the 3-$\sigma$ level. Using the new constraints provided by the H-band data (including both high-accuracy visibilities and closure phases) and a significantly improved spatial coverage, the most straightforward scenario consists in a compact dust disc producing a thermal emission that is largely dominated by small grains located between 0.1 and 0.3\,AU from Vega and accounting for a relative flux with respect to the stellar photosphere of 1.23 $\pm$ 0.45\%. This flux ratio is shown to vary slightly with the geometry of the model used to fit our interferometric data (variations within $\pm$0.19\%). Using the new H-band flux ratio together with archival measurements at various wavelengths, we show by means of SED modelling that at least a small fraction of carbonaceous grains must be present in the disc in order to fit the data in a satisfactory way. A steep density profile is also necessary to ensure the compatibility with the non-detection reported by nulling interferometry in the mid-infrared. Considering the best-fit model, a dust mass of approximately $2\times10^{-9}M_\oplus$, equivalent to the mass of an asteroid about 20\,km in diameter, would be necessary to explain the near-infrared excess emission. Given the short lifetime of dust in the inner Vega system, a major dynamical event, similar to the solar system's late heavy bombardment, might be currently ongoing in the Vega system. This would support the idea that the debris disc around Vega is the only visible component of a more complex planetary system harbouring unknown planets.

\begin{acknowledgements}
The authors acknowledge Sylvestre Lacour (LESIA) and Ettore Pedretti (St.\,Andrews) for sharing dispersed calibrator data of 2006. The authors are also grateful to Arnaud Magette, Charles Hanot, Pierre Riaud, and Jean Surdej (IAGL), Bertrand Mennesson (NASA/JPL), Jason Aufdenberg (ERAU), Gerd Weigelt (MPIFR), and Paul Lepoulpe for helpful advice. This research was supported by the International Space Science Institute (ISSI) in Bern, Switzerland (``Exozodiacal Dust discs and Darwin" working group, \textit{http://www.issibern.ch/teams/exodust/}). DD acknowledges the support of the Belgian National Science Foundation (``FRIA"), of EII (Fizeau programme), and the MPIFR. OA acknowledges the support from an F.R.S.-FNRS Postdoctoral Fellowship. DD and OA acknowledge support from the Communaut\'e fran\c{c}aise de Belgique - Actions de recherche concert\'ees - Acad\'emie universitaire Wallonie-Europe. DD, OA, and JCA thank the French National Research Agency (ANR) for financial support through contract ANR-2010 BLAN-0505-01 (EXOZODI). This research received the support of PHASE, the high angular resolution partnership between ONERA, Observatoire de Paris, CNRS and University Denis Diderot Paris 7.
\end{acknowledgements}

\bibliography{C:/Users/Den/Work/biblio}
\bibliographystyle{aa}

\end{document}